\documentclass[aps,prl, twocolumn,superscriptaddress,nofootinbib,preprintnumbers]{revtex4-1}

\usepackage{graphicx,amsfonts,amssymb,amsmath,color}
\usepackage[colorlinks,citecolor=red]{hyperref}
\allowdisplaybreaks[4]
\begin{document}
\preprint{
{\vbox {
\hbox{\bf MSUHEP-18-007}
}}}
\vspace*{0.2cm}

\newcommand{\beq}{\begin {equation}}
\newcommand{\eeq}{\end   {equation}}
\newcommand{\bea}{\begin {eqnarray}}
\newcommand{\eea}{\end   {eqnarray}}
\newcommand{\nn }{\nonumber        }

\newcommand{\red}{\textcolor{red}}

\title{Signature of Pseudo Nambu-Goldstone Higgs boson in its Decay}

\author{Qing-Hong Cao}
\email{qinghongcao@pku.edu.cn}
\affiliation{Department of Physics and State Key Laboratory of Nuclear Physics and Technology, Peking University, Beijing 100871, China}
\affiliation{Collaborative Innovation Center of Quantum Matter, Beijing, 100871, China}
\affiliation{Center for High Energy Physics, Peking University, Beijing 100871, China}

\author{Ling-Xiao Xu}
\email{lingxiaoxu@pku.edu.cn}
\affiliation{Department of Physics and State Key Laboratory of Nuclear Physics and Technology, Peking University, Beijing 100871, China}

\author{Bin Yan}
\email{yanbin1@msu.edu}
\affiliation{Department of Physics and Astronomy, Michigan State University, East Lansing, MI 48824, USA}

\author{Shou-hua Zhu}
\email{shzhu@pku.edu.cn}
\affiliation{Department of Physics and State Key Laboratory of Nuclear Physics and Technology, Peking University, Beijing 100871, China}
\affiliation{Collaborative Innovation Center of Quantum Matter, Beijing, 100871, China}
\affiliation{Center for High Energy Physics, Peking University, Beijing 100871, China}

\begin{abstract}
If the Higgs boson is a pseudo Nambu-Goldstone boson (PNGB), the $hZ\gamma$ contact interaction induced by the $\mathcal{O}(p^4)$ invariants of the non-linear sigma model is free from its nonlinearity effects. The process $h\rightarrow Z\gamma$ can be used to eliminate the universal effects of heavy particles, which can fake the nonlinearity effects of the PNGB Higgs boson in the process $h\rightarrow V^*V$ ($V=W^\pm$,\ $Z$). We demonstrate that the ratio of the signal strength of $h\rightarrow Z\gamma$ and $h\rightarrow V^*V$ is good to distinguish the signature of the PNGB Higgs boson from Higgs coupling deviations.
\end{abstract}

\maketitle

\noindent{\bf Introduction.} Deciphering the nature of the Higgs boson is one of the major tasks of particle physics, and one can ask whether there is dynamics behind electroweak symmetry breaking (EWSB). Given no hint of new heavy particles at the Large Hadron Collider (LHC), the best strategy is to use effective theories to parametrize the ignorance of UV physics. When the Higgs boson arises from a weakly-coupled UV theory, the Standard Model (SM) Effective Field Theory (EFT) can be used. On the other hand, the Higgs boson might emerge  as a pseudo Nambu-Goldstone boson (PNGB)  from some strong dynamics at the TeV scale~\cite{Kaplan:1983fs, Kaplan:1983sm, Dugan:1984hq, ArkaniHamed:2001nc, ArkaniHamed:2002qy, Contino:2003ve, Agashe:2004rs,Chacko:2005pe}; see Refs.~\cite{Contino:2010rs, Bellazzini:2014yua, Panico:2015jxa} for recent reviews. 
The traditional CCWZ formalism~\cite{Coleman:1969sm, Callan:1969sn}  is often used to construct the non-linear sigma model (NL$\sigma$M) of the PNGB Higgs boson with the symmetry breaking pattern $\mathcal{G}/\mathcal{H}$. Alternatively, one can use the so-called shift symmetry~\cite{Low:2014nga,Low:2014oga} to construct NL$\sigma$M even without knowing the UV group $\mathcal{G}$. The nature of the PNGB Higgs boson is encoded in the NL$\sigma$M, and one can characterize its signature explicitly with a parameter $\xi$, which is defined as the ratio of the electroweak scale $v$ and the decay constant of the PNGB Higgs boson $f$.
For that the parameter $\xi$ is named as the {\it{nonlinearity}} of the PNGB Higgs boson.

It's crucial to tell whether the Higgs boson is a PNGB from Higgs  precision measurements. In particular, the Higgs couplings to electroweak gauge bosons are of the most importance as they are  directly related to the EWSB. Unfortunately, one cannot learn any useful information of the parameter $\xi$ from  the $hVV$ ($V=W^\pm,Z$) couplings  alone. For example,  two effects could  modify  the $hVV$ couplings and fake each other:
\begin{enumerate}
  \item  the nonlinearity of the PNGB Higgs boson $\xi$;
  \item  the shift-symmetry-breaking effects induced by heavy particles, e.g.  a singlet scalar interacting with the Higgs boson~\cite{Low:2009di,Dawson:2017vgm,Corbett:2017ieo}. 
\end{enumerate}
To probe the nonlinearity of the PNGB Higgs boson, we propose an observable $R$ defined as the ratio of the signal strengths of the $h\rightarrow Z\gamma$ and $h\rightarrow V^*V$ decay channels,
\bea
R\equiv \frac{\mu(h\rightarrow Z\gamma )}{\mu (h\rightarrow V^*V )}.
\eea 
We demonstrate that the ratio $R$ is sensitive only to nonlinearity of the PNGB Higgs boson in strongly-coupled models but not to the faking effects originating from unknown heavy particles. 
Another advantage is that $R$ is independent of the single Higgs  boson cross section and the Higgs boson width.

\noindent{\bf Higgs Couplings to Electroweak Gauge Bosons.} With only the information of the group $\mathcal{H}$ in the IR, the NL$\sigma$M of the PNGB Higgs boson can be constructed with the shift symmetry covariants $\widetilde{D}_\mu H$ and $\mathcal{E}_\mu$~\cite{Low:2014nga,Low:2014oga}. $\widetilde{D}_\mu H$ is the so-called Goldstone covariant derivative and $\mathcal{E}_\mu$ is the associated gauge fields of the shift symmetry. The constructed NL$\sigma$M is universal, up to the normalization of the PNGB decay constant $f$,  as the PNGB Higgs  boson can be embedded in different UV group $\mathcal{G}$.
In this work, we focus on the case that $\mathcal{H}$ contains the custodial $SO(4)$ symmetry and the PNGB Higgs boson arises from a custodial $4$-plet. Below the scale of heavy resonances of the strong dynamics, the NL$\sigma$M can be expanded as
\bea
\mathcal{L}_{\textnormal{NL$\sigma$M}}=\mathcal{O}(p^2)+ \mathcal{O}(p^4)+\cdots\ .
\eea
From the operators at the order of $\mathcal{O}(p^2)$ and $\mathcal{O}(p^4)$ of the NL$\sigma$M, one can obtain the Higgs couplings to electroweak gauge bosons which respect the shift symmetry of the PNGB Higgs boson.

At the order of $\mathcal{O}(p^2)$, there is only one invariant $(\widetilde{D}_\mu H)^\dagger \widetilde{D}^\mu H$, which gives rise to both the normalized kinetic term of the physical Higgs boson and the mass term of the electroweak gauge bosons, i.e.   
\bea
&&(\widetilde{D}_\mu H)^\dagger \widetilde{D}^\mu H =\frac{1}{2}\partial_\mu h\partial^\mu h\nn\\
&&\ \ +(2f^2) \frac{g^2}{4}\sin^2\frac{\langle h\rangle+h}{\sqrt{2} f}\left(W_\mu^+W^{-\mu}+\frac{Z^\mu Z_\mu}{2 \cos^2\theta_W}\right),
\eea
where $\theta_W$ is the weak mixing angle and $g$ is the gauge coupling of $SU(2)_L$ group. Matching the $W^\pm$ and $Z$ masses yields the electroweak VEV as 
\bea
v=\sqrt{2}f\sin\frac{\langle h\rangle}{\sqrt{2} f}=246\ \textnormal{GeV}.
\label{eq:vev}
\eea
Note that the VEV of the physical Higgs boson, $\langle h\rangle$, is not exactly equal to $246$ GeV. We  define the nonlinearity parameter $\xi$ as
\bea
\xi\equiv \frac{v^2}{2f^2}=\sin^2\frac{\langle h\rangle}{\sqrt{2} f}\ .
\eea
 The $hVV$ coupling is
 \bea
\mathcal{L}_{hVV}=\frac{M_V^2}{v}\sqrt{1-\xi}\ h V_\mu V^\mu,
\label{eq:hvv}
\eea
where $M_V$ is the mass of electroweak gauge boson $V$. 

Although sensitive to $\xi$,  the $hVV$ coupling alone cannot provide enough information to pin down the Higgs boson nature;  the $hVV$ coupling could be modified by  heavy particles which violate the shift symmetry~\cite{Low:2009di,Dawson:2017vgm,Corbett:2017ieo}.
We use SM-EFT to describe the new physics (NP) effects which violate the shift symmetry in Higgs boson physics. 
Only one leading operator needs to be considered at dimension-six level~\cite{Giudice:2007fh},
\begin{equation}
O_H=\frac{1}{2v^2}\partial_\mu(H^\dagger H)\partial^\mu(H^\dagger H).
\end{equation}
The effect of $O_H$ is universal in all the single Higgs boson processes as it simply rescales the amplitude as $h\to h/\sqrt{1+c_H}$ due to the renormalization of the Higgs boson field. For cancelling the universal $O_H$ effect, we further consider the $hZ\gamma$ coupling.

Within the NL$\sigma$M, the leading contribution to the $hZ\gamma$ effective coupling arises from the order of $\mathcal{O}(p^4)$. All the relevant $\mathcal{O}(p^4)$ operators can be derived in the CCWZ formalism with the coset $SO(5)/SO(4)$~\cite{Contino:2011np,Azatov:2013ura}. Alternatively, one can derive the $\mathcal{O}(p^4)$ operators based on the shift symmetry~\cite{Liu:2018vel,Liu:2018qtb}. The resultant operators are valid 
in any other cosets as long as there is an unbroken $SO(4)$ symmetry in the IR with a Higgs boson $4$-plet. 
The $hZ\gamma$ effective coupling arises from two operators, $\widetilde{O}_{HB}=ig^\prime(\widetilde{D}^\mu H)^\dagger(\widetilde{D}^\nu H)B_{\mu\nu}$ and $\widetilde{O}_{HW}=ig(\widetilde{D}^\mu H)^\dagger\sigma^i(\widetilde{D}^\nu H)W^i_{\mu\nu}$, which are shown as follows,
\begin{align}
&ig^\prime(\widetilde{D}^\mu H)^\dagger(\widetilde{D}^\nu H)B_{\mu\nu}\nn\\
&=-\frac{gg^\prime}{4\cos\theta_W}(\sqrt{2}f)\sin\frac{h+\langle h\rangle}{\sqrt{2}f}(\partial^\mu h Z^\nu-\partial^\nu h Z^\mu)B_{\mu\nu}\nn\\
&\ \ \ \ \ +\cdots,\\
&ig(\widetilde{D}^\mu H)^\dagger\sigma^i(\widetilde{D}^\nu H)W^i_{\mu\nu}\nn\\
&=\frac{g^2}{4\cos\theta_W}(\sqrt{2}f)\sin\frac{h+\langle h\rangle}{\sqrt{2}f}(\partial^\mu hZ^\nu-\partial^\nu h Z^\mu)\nn\\
&\ \ \ \ \times(\partial_\mu W^3_\nu-\partial_\nu W^3_\mu)+\cdots,
\end{align}
where $B_{\mu\nu}$ and $W^3_{\mu\nu}$ are the field strength tensors of external electroweak gauge bosons and $g^\prime$ is the gauge coupling of the  $U(1)_Y$ group. The $hZ\gamma$  coupling is 
\begin{align}
\mathcal{L}_{hZ\gamma}&=(\widetilde{c}_{HW}\widetilde{O}_{HW}+\widetilde{c}_{HB}\widetilde{O}_{HB})/M_W^2\nn\\
&=-\Delta\kappa_{Z\gamma} \tan{\theta_W}\frac{1}{v}(\partial^\mu h Z^\nu-\partial^\nu h Z^\mu) A_{\mu\nu},
\end{align} 
with  $\Delta\kappa_{Z\gamma}=\widetilde{c}_{HB}-\widetilde{c}_{HW}$.
The $hZ\gamma$ coupling induced by the $\mathcal{O}(p^4)$ invariants of the NL$\sigma$M does not depend on $\xi$, i.e. it is not sensitive to the nonlinearity of the PNGB Higgs boson at all.

Both operators $\widetilde{O}_{HW}$ and $\widetilde{O}_{HB}$ modify the $hVV$ and $hZ\gamma$ couplings. However, we only consider their effects in the $hZ\gamma$ coupling due to the reasons as follows. The ratio $R$ of interest to us is
\begin{equation}
R =  \dfrac{\mu(h\to Z\gamma)}{\mu(h\to V^*V)}\propto \dfrac{g_{hZ\gamma}}{g_{hZ\gamma}^{\rm SM}} \cdot\dfrac{g_{hVV}^{\rm SM}}{g_{hVV}},
\end{equation}
where $g_{hZ\gamma}^{\rm SM}$ ($g_{hZ\gamma}$) and $g_{hVV}^{\rm SM}$ ($g_{hVV}$) denotes the  couplings of $hZ\gamma$ and $hVV$ in the SM (NP), respectively. The $hVV$ coupling is generated at tree level while the $hZ\gamma$ coupling at one loop level in the SM. To match the same precision of the SM couplings, the sub-leading correction to $hVV$ coupling from $\widetilde{O}_{HW}$ and $\widetilde{O}_{HB}$ are neglected due to the loop suppression~\cite{Giudice:2007fh}. Similarly all the operators are considered in the $h\rightarrow Z\gamma$ decay.

There is another loop-induced and shift-symmetry breaking operator $O_\gamma\sim H^\dagger HB_{\mu\nu}B^{\mu\nu}$ contributing to the processes of $h\to Z\gamma$ and $h\to \gamma\gamma$. Its contribution is tightly constrained by the precision measurements of $h\gamma\gamma$ coupling, however. We thus neglect the contribution of $O_\gamma$ in $h\rightarrow Z\gamma$ in this work.

 \noindent{\bf The Ratio $R$.} Next we show the observable $R$, the ratio of the signal strengths of $h\rightarrow Z\gamma$ and $h\rightarrow V^* V$,
is sensitive only to the nonlinearity of the PNGB Higgs boson but not to the faking effects of $O_H$. 

The leading corrections to the signal strength of the process $h\rightarrow V^* V$ from both the nonlinearity of the PNGB Higgs boson  and the $O_H$ operator are
\bea
\mu(h\rightarrow V^* V)&=&\frac{\sigma_{h}\times\textnormal{BR}(h\rightarrow V^* V)}{\sigma_{h}^{\textnormal{SM}}\times\textnormal{BR}(h\rightarrow V^* V)_{\textnormal{SM}}}\nn\\
&=&\frac{\sigma_{h}}{\sigma^{\textnormal{SM}}_{h}}\cdot\frac{\Gamma^{\textnormal{SM}}_{\textnormal{total}}}{\Gamma_\textnormal{total}}\cdot F_{\rm PNGB}\cdot F_{ O_H}\ ,
\label{hvv}
\eea
where 
\beq
F_{\rm PNGB} =1-\xi, \quad F_{O_H}=\frac{1}{1+c_H}.
\eeq
Here, $\sigma_{h}$  ($\sigma_{h}^{\rm SM}$) and $\Gamma_{\rm {total}}$ ($\Gamma_{\rm {total}}^{\rm SM}$) denotes the single Higgs boson production cross section and the Higgs boson  total width in  NP models (the SM), respectively.
The substitution of $\xi\rightarrow -\xi$ describes a PNGB Higgs boson from various non-compact groups~\cite{Low:2014oga, Alonso:2016btr}.  It is clear that one cannot distinguish the contribution from $F_{\rm PNGB}$ and $F_{O_H}$ in the $h\to VV^*$ decays.

\begin{figure}
\includegraphics[scale=0.21]{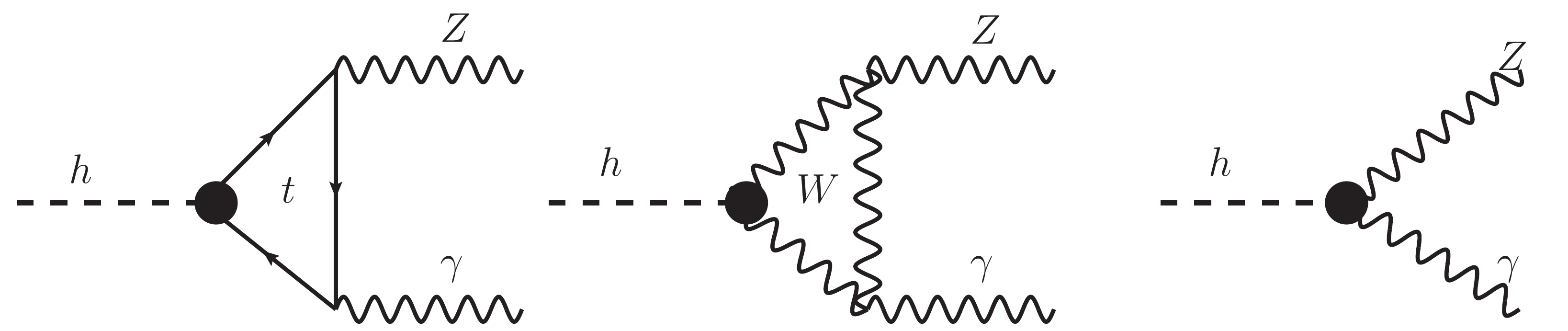}
\caption{Illustrative Feynman diagrams of $h\to Z\gamma$. The black dots denote the effective couplings including both the SM and NP effects. }
\label{fig:feyman}
\end{figure}

\begin{table}
\caption{The NP effect in the $hVV$ and $hZ\gamma$ couplings when the Higgs boson is a PNGB or a SM-like scalar ($\xi\to 0$) from weakly-coupled UV theories. The symbol $\surd$ and $\huge\times$ means that the NP effect could and cannot contribute, respectively.}
\begin{center}
\begin{tabular}{r|c|c|r|c|c}
\hline\hline
$hVV$ &PNGB& SM-like& $hZ\gamma$ & PNGB & SM-like\\
\hline
$\xi$-effect &$ \surd$&$\huge\times$& ~~$\xi$-effect &$\huge\times$ &$\huge\times$\\
\hline
$O_H$&$\surd$ &$\surd$ & $O_H$&$\surd$ &$\surd$\\
\hline\hline
\end{tabular}
\end{center}
\label{tbl:hvv}
\end{table} 

For the process of $h\rightarrow Z\gamma$ there are contributions from the top-quark loop, the $W$-boson loops and the $hZ\gamma$ effective coupling; see Fig.~\ref{fig:feyman}. NP effects that modify the $hWW$ and $ht\bar{t}$ effective couplings are also included. The partial width of the $h\rightarrow Z\gamma$ decay is
\bea
\Gamma(h\to Z\gamma)&=&F_{O_H}\ \dfrac{M_h^3}{8\pi v^2}\left(1-\dfrac{M_Z^2}{M_h^2}\right)^3\nn\\
&\times&\left|F_{Z\gamma}^{t}+F_{Z\gamma}^{W}\sqrt{F_{\rm PNGB}}+\Delta\kappa_{Z\gamma}\tan\theta_W\right|^2,\qquad
\eea
where $M_h$ and $M_Z$ denotes the masses of the Higgs boson and the $Z$ boson, respectively. $F_{Z\gamma}^W$ and $F_{Z\gamma}^t$ represents the loop effects of the $W$-boson and top-quark in the SM. Note that the $W$-boson loop dominates over the top-quark loop, e.g. $F_{Z\gamma}^W=0.0087$ and $F_{Z\gamma}^t=-0.00097$. 
The signal strength of $h\to Z\gamma$ is 
\bea
\mu(h\to Z\gamma)&=&\dfrac{\sigma_h\times {\rm BR}(h\to Z\gamma)}{\sigma_h^{\rm SM}\times {\rm BR}(h\to Z\gamma)_{\rm SM}}\nn\\
&=&\frac{\sigma_h}{\sigma^{\textnormal{SM}}_h}\cdot\frac{\Gamma^{\textnormal{SM}}_{\textnormal{total}}}{\Gamma_\textnormal{total}}\cdot F_{O_H}\nn\\
&\times& \dfrac{\left|F_{Z\gamma}^{t}+F_{Z\gamma}^{W}\sqrt{F_{\rm PNGB}}+\Delta\kappa_{Z\gamma}\tan\theta_W\right|^2}{|F_{Z\gamma}^{t}+F_{Z\gamma}^{W}|^2},\qquad
\label{hza}
\eea
where the $O_H$ effect ($F_{OH}$) is factorized out.

The ratio $R$ follows from Eqs.~(\ref{hvv}) and~(\ref{hza}) as 
\bea
R=\dfrac{\left|F_{Z\gamma}^{t}+F_{Z\gamma}^{W}\sqrt{F_{\rm PNGB}}+\Delta\kappa_{Z\gamma}\tan\theta_W\right|^2}{|F_{Z\gamma}^{t}+F_{Z\gamma}^{W}|^2\ F_{\rm PNGB}}.
\label{eq:R}
\eea
The dependence of $\sigma_h$ and $\Gamma_\textnormal{total}$ cancels out in the ratio $R$, and, more important, the $F_{O_H}$ term also cancels out.
Table~\ref{tbl:hvv} shows the impact of both the Higgs nonlinearity and the $O_H$ operator on the $hVV$ and $hZ\gamma$ effective couplings, depending on whether the Higgs boson is a PNGB  from strong dynamics at $\sim{\rm TeV}$ scale or a SM-like scalar  from  weakly-coupled UV theories ($\xi\to 0$). The $hZ\gamma$ effective coupling is crucial to eliminate the universal effect of $O_H$ so as to extract out the nonlinearity ($\xi$) of the PNGB Higgs boson.

One thus can determine $F_{\rm PNGB}$ when both the $\Delta\kappa_{Z\gamma}$ and $R$ are known precisely from data; for example, $F_{\rm PNGB}$ follows directly from Eq.~\ref{eq:R} as
\beq
F_{\rm PNGB}=\left(\dfrac{F_{Z\gamma}^{t}+\Delta\kappa_{Z\gamma}\tan\theta_W}{\sqrt{R}|F_{Z\gamma}^{t}+F_{Z\gamma}^{W}|-F_{Z\gamma}^{W}}\right)^2 \simeq\left(\dfrac{\Delta\kappa_{Z\gamma}\tan\theta_W}{(\sqrt{R}-1)F_{Z\gamma}^W}\right)^2
\label{eq:FPNGB}
\eeq
when $\sqrt{F_{\rm PNGB}} > 0.11-0.6\times\left(\tfrac{\Delta\kappa_{Z\gamma}}{0.01}\right)$, or as
\beq
F_{\rm PNGB}=\left(\dfrac{F_{Z\gamma}^{t}+\Delta\kappa_{Z\gamma}\tan\theta_W}{\sqrt{R}|F_{Z\gamma}^{t}+F_{Z\gamma}^{W}|+F_{Z\gamma}^{W}}\right)^2 \simeq\left(\dfrac{\Delta\kappa_{Z\gamma}\tan\theta_W}{(\sqrt{R}+1)F_{Z\gamma}^W}\right)^2
\label{eq:FPNGB2}
\eeq
when $\sqrt{F_{\rm PNGB}} <0.11-0.6\times\left(\frac{\Delta\kappa_{Z\gamma}}{0.01}\right)$,
where the approximations can be understood by $F_{Z\gamma}^W\gg |F_{Z\gamma}^t|$.
For a sizable $\xi$ one might be able to tell the PNGB Higgs boson apart from a SM-like scalar (i.e. $F_{\rm PNGB}= 1$). If the Higgs boson is a PNGB, $F_{\rm PNGB}$ could be smaller than one or larger than one, depending on a specific UV group from which the PNGB Higgs boson emerges; for example, $F_{\rm PNGB}< 1$ for a compact UV group ($\xi>0$) and $F_{\rm PNGB}> 1$ for a non-compact UV group ($\xi<0$)~\cite{Alonso:2016btr}. 
It is fascinating that the ratio $R$ distinguishes the compactness of the broken UV-group $\mathcal{G}$'s.

\begin{figure}[b!]
\includegraphics[scale=0.23]{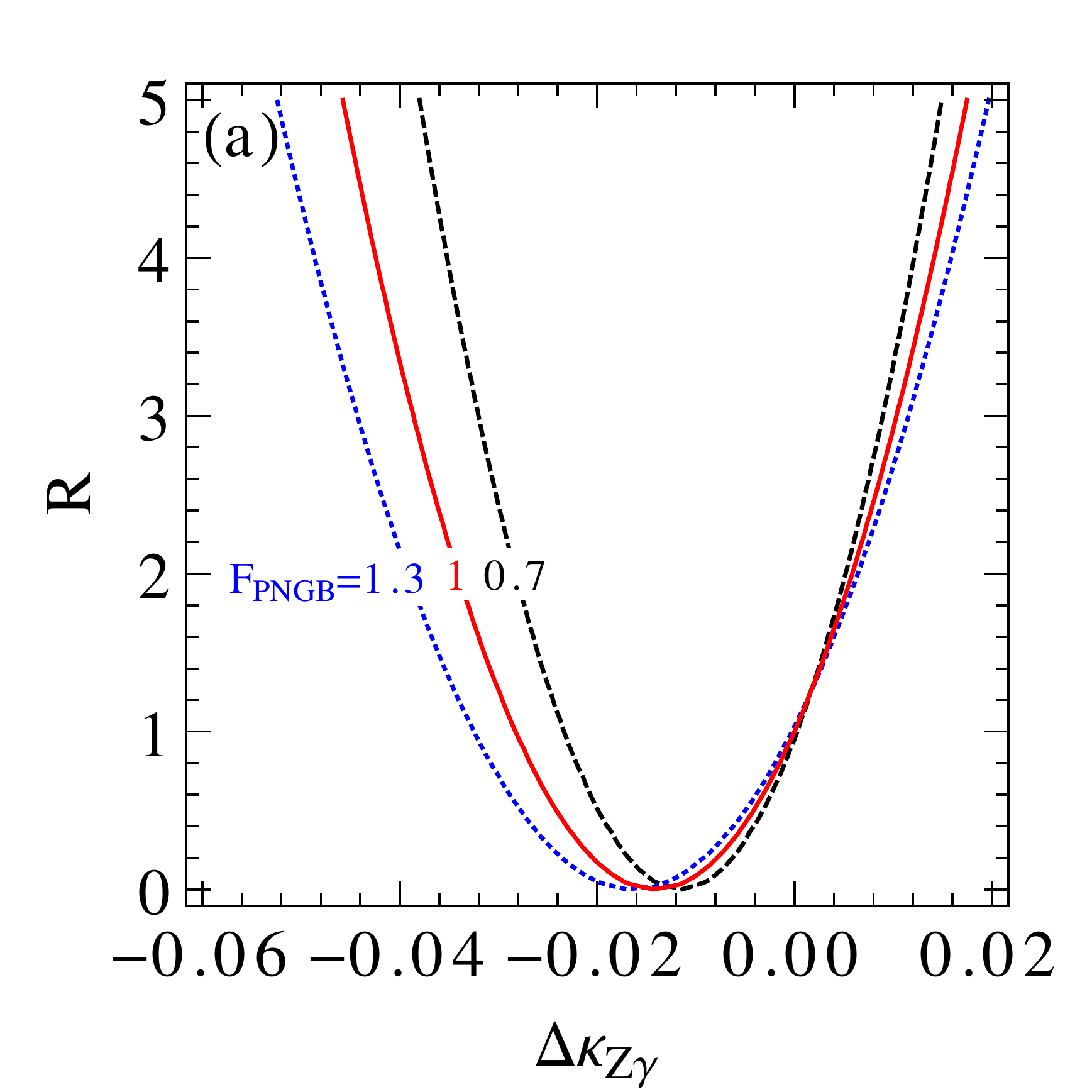}
\includegraphics[scale=0.23]{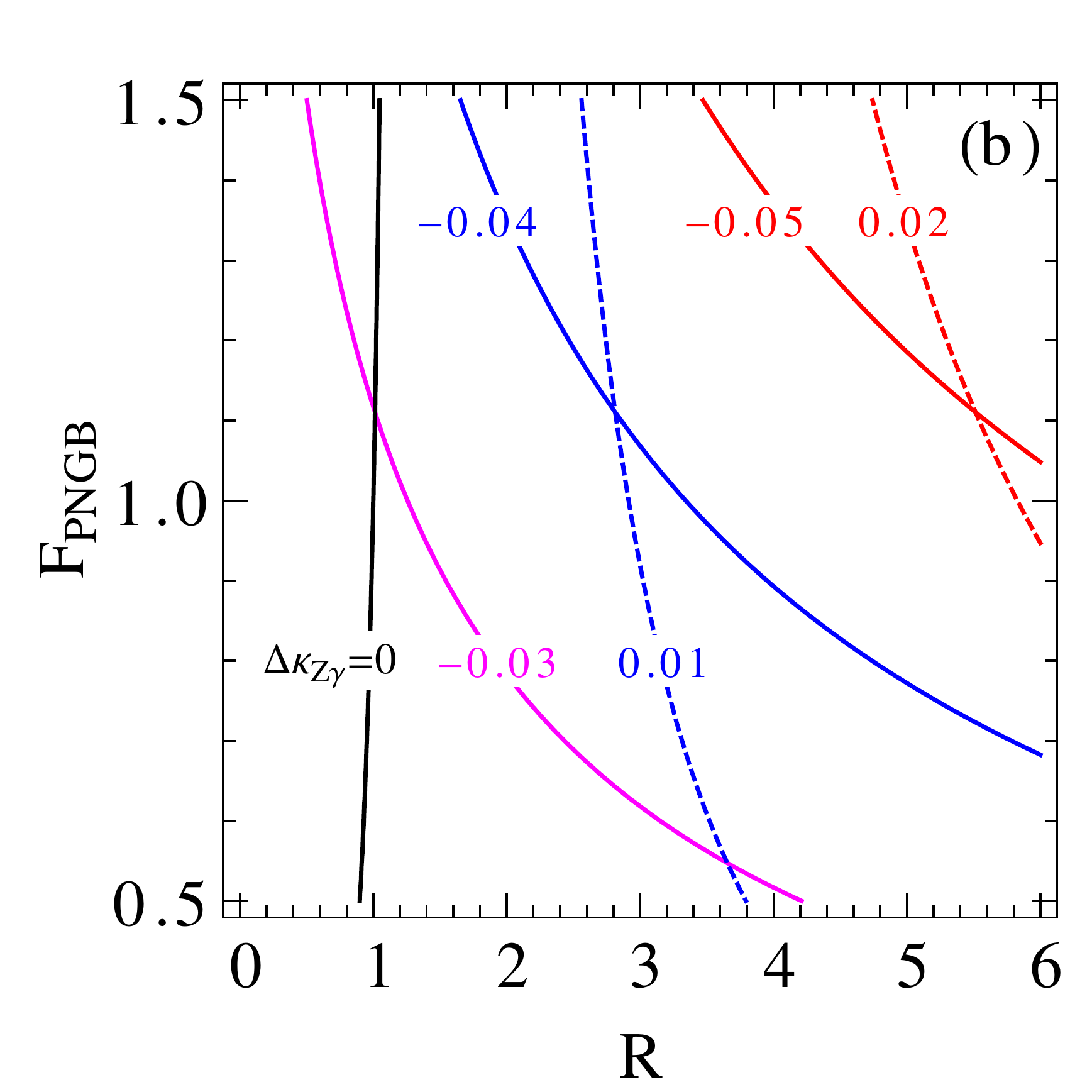}
\caption{The correlations among $F_{\rm PNGB}$, $R$ and $\Delta \kappa_{Z\gamma}$. Shown is the contour line of one parameters in the plane of the other twos. }
\label{fig0}
\end{figure}

Figure~\ref{fig0}(a) displays the contour of $F_{\rm PNGB}$ in the plane of $R$ and $\Delta \kappa_{Z\gamma}$ for $F_{\rm PNGB}=0.7$ (black), 1.0 (red) and 1.3 (blue). Figure~\ref{fig0}(b) shows the dependence of $F_{\rm PNBG}$ on $R$ for various $\Delta\kappa_{Z\gamma}$'s. We note that the discrimination power of $F_{\rm PNGB}$ increases with $R$ and is strong for a negative $\Delta\kappa_{Z\gamma}$ but quite weak for positive $\Delta\kappa_{Z\gamma}$'s. For example, the blue dashed curve ($\Delta \kappa_{Z\gamma}=0.01$) in Fig.~\ref{fig0}(b) is not sensitive to $F_{PNGB}$.

\noindent{\bf Sensitivity at the LHC and CEPC.~}  Now consider the potential of measuring $F_{\rm PNGB}$ at the High luminosity Large Hadron Collider (HL-LHC), a proton-proton collider to operate at $E_{\rm cm}=14~{\rm TeV}$ with an integrated luminosity of $3~{\rm ab}^{-1}$~\cite{Apollinari:2116337} , and also at the Circular electron-positron collider (CEPC), proposed to operator at $E_{\rm cm}=240~{\rm GeV}$ with an integrated luminosity of $5~{\rm ab}^{-1}$~\cite{CDR}. 

The coefficient $\Delta\kappa_{Z\gamma}$ can be derived from the measurement of anomalous triple gauge-boson couplings (aTGCs)~\cite{DeRujula:1991ufe, Hagiwara:1993ck}
\bea
{\cal L}_{\rm TGC}/g_{WW\mathbb{V}} &=&
ig_{1,\mathbb{V}} \Big( W^+_{\mu\nu}W^-_{\mu}\mathbb{V}_{\nu} -W^-_{\mu\nu}W^+_{\mu}\mathbb{V}_{\nu} \Big)\nn\\
&+& i\kappa_{\mathbb{V}} W^+_\mu W^-_\nu \mathbb{V}_{\mu\nu}+ \frac{i\lambda_{\mathbb{V}}}{M_W^2} W^+_{\lambda\mu} W^-_{\mu\nu} \mathbb{V}_{\nu\lambda},\qquad
\eea
where $\mathbb{V}=\gamma/Z$, $g_{WW\gamma}=-e$ and $g_{WWZ}=-e\cot\theta_W$. The NP contributions in the $g_{1,Z}$ and $\Delta\kappa_{\gamma}$ are 
\bea
\Delta g_{1,Z}&\equiv&g_{1,Z}-1=\widetilde{c}_{HW}/\cos^2\theta_W,\nn\\
\Delta \kappa_\gamma &\equiv&\kappa_{\gamma}-1=  \widetilde{c}_{HW} + \widetilde{c}_{HB}.
\eea
It follows that
\beq
\Delta\kappa_{Z\gamma}=\widetilde{c}_{HB}-\widetilde{c}_{HW}=\Delta\kappa_\gamma-2\Delta g_{1,Z}\cos\theta_W^2.
\eeq
$\Delta \kappa_\gamma$ and $\Delta g_{1,Z}$ are expected to be measured with precisions as follows~\cite{Bian:2015zha}:
\begin{align}
&& \delta\kappa_\gamma &=& 0.0029~(\text{HL-LHC}), &&0.00022~(\text{CEPC}),\nn\\
&& \delta g_{1,Z}  &=& 0.0011~(\text{HL-LHC}), && 0.00016~(\text{CEPC}).
\end{align}
The uncertainty $\delta\kappa_{Z\gamma}$, 
\bea
\delta\kappa_{Z\gamma}=\sqrt{\delta\kappa_\gamma^2+4\cos\theta_W^4\delta g_{1,Z}^2},\nn
\eea
is given by
\begin{align}
&& \delta\kappa_{Z\gamma} &=& 0.0033~(\text{HL-LHC}), &&0.00034~(\text{CEPC}).
\label{eq:kza_error}
\end{align}
The uncertainty of the ratio $R$ is 
\bea
\dfrac{\delta R}{R_0}=\sqrt{\left(\dfrac{\delta \mu_{h\to Z\gamma}}{\mu_{h\to Z\gamma}^0}\right)^2+\left(\dfrac{\delta \mu_{h\to VV^*}}{\mu_{h\to VV^*}^0}\right)^2},
\label{eq:err}
\eea
where $R_0$, $\mu_{h\to Z\gamma}^0$ and $\mu_{h\to VV^*}^0$ denotes the central values of $R$, $\mu(h\to Z\gamma)$ and $\mu(h\to VV^*)$, respectively. The signal strengths $\mu(h\to Z\gamma)$ and $\mu(h\to VV^*)$ are expected to be measured at the HL-LHC~\cite{ATL-PHYS-PUB-2014-006,CMS-NOTE-13-002}  and the CEPC~\cite{Durieux:2017rsg} with errors as follows: 
\begin{align}
&& \delta\mu_{h\to Z\gamma} &=& 0.3~(\text{HL-LHC}), &&0.25~(\text{CEPC}),\nn\\
&& \delta\mu_{h\to VV^*}&=& 0.1~(\text{HL-LHC}), && 0.01~(\text{CEPC}).
\end{align}
For simplicity we assume $\mu_{h\to VV^*}^0=1$ (i.e. no deviation in the $hVV$ couplings), yielding  
\bea
\delta R=\sqrt{\left(\delta \mu_{h\to Z\gamma}\right)^2+R_0^2\left(\delta \mu_{h\to VV^*}\right)^2}.
\label{err_h}
\eea
Defining $F_{\rm PNGB}^0$ and $\Delta\kappa_{Z\gamma}^0$ as the central values of $F_{\rm PNGB}$ and $\Delta\kappa_{Z\gamma}$, respectively, the error of $F_{\rm PNGB}$ is 
\begin{align}
&\dfrac{\delta\sqrt{F_{\rm PNGB}}}{\sqrt{F_{\rm PNGB}^0}}=\sqrt{\Big(\dfrac{\delta\kappa_{Z\gamma}}{\Delta\kappa_{Z\gamma}^0}\Big)^2+\Big(\dfrac{\delta\sqrt{R}}{\sqrt{R_0} \mp 1}\Big)^2}\nn\\
&\simeq 0.32\times \sqrt{\Big(\dfrac{\delta\kappa_{Z\gamma}}{0.003}\Big)^2
\Big(\dfrac{0.01}{\Delta\kappa_{Z\gamma}^0}\Big)^2+\Big(\dfrac{\delta R}{0.3}\Big)^2\dfrac{1}{4R_0(\sqrt{R_0} \mp 1)^2}},
\label{eq:errF1}
\end{align}
where we normalize the errors with the HL-LHC projections and the sign ``$\mp$" refers to Eqs.~\ref{eq:FPNGB} and \ref{eq:FPNGB2}. A large $R_0$ would suppress the effects $\delta R$ and we expect to reach a better measurement of $F_{\rm PNGB}$.  
The typical error of $F_{\rm PNGB}$ is about 30\% at the HL-LHC, while it reduces to $\sim3\%$ at the CEPC.

\begin{figure}[b]
\includegraphics[scale=0.242]{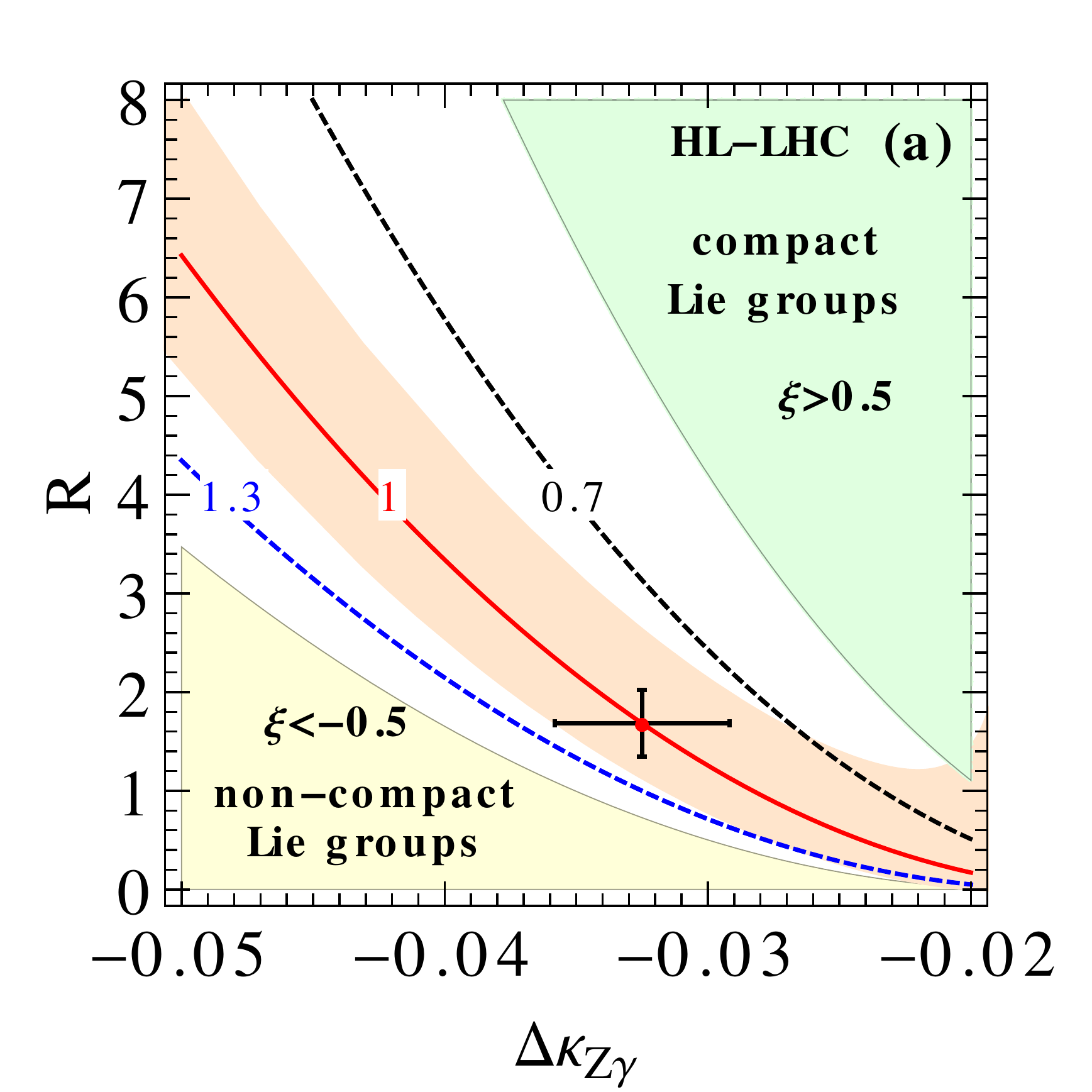}
\includegraphics[scale=0.242]{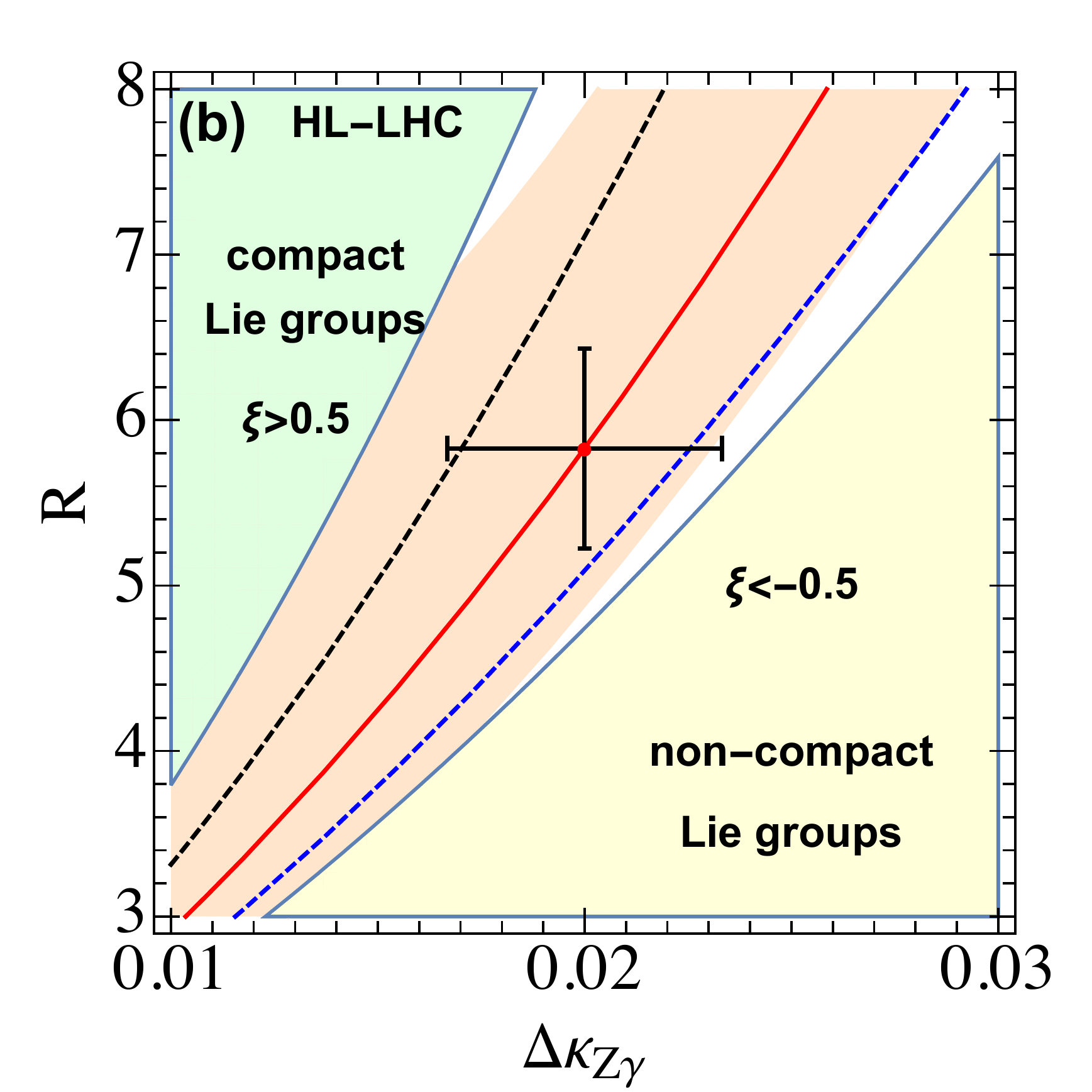}
\includegraphics[scale=0.242]{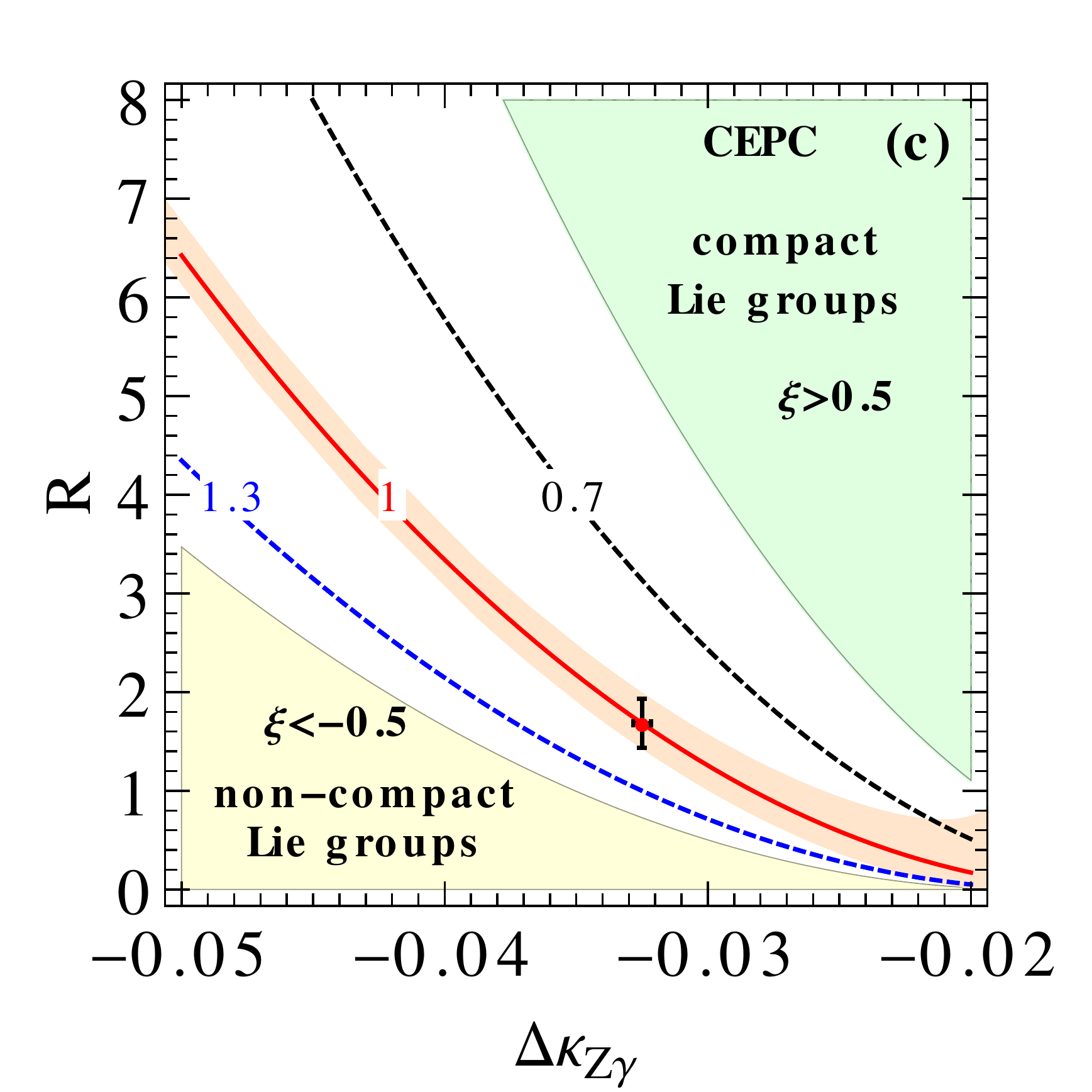}
\includegraphics[scale=0.242]{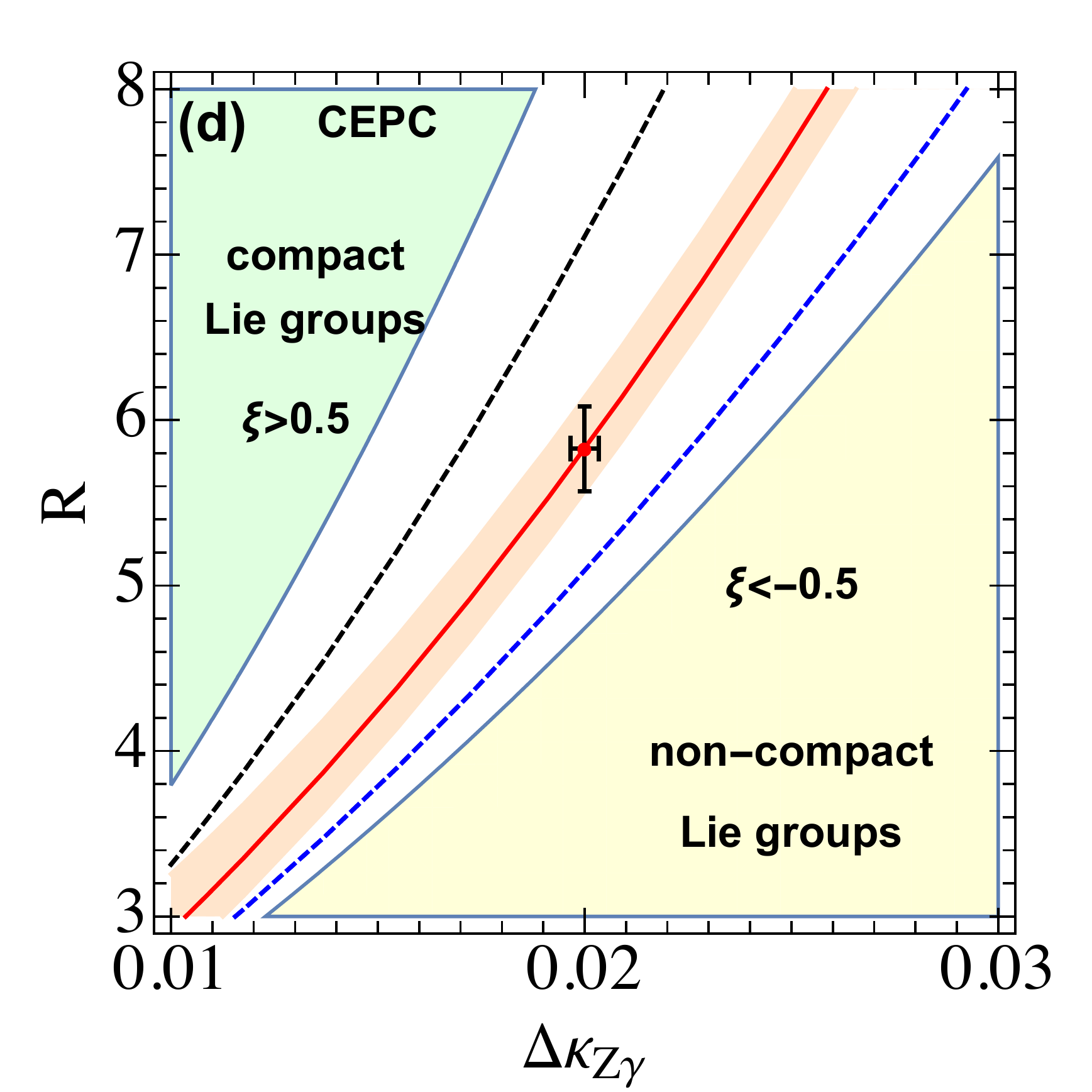}
\caption{The sensitivity to $F_{\rm PNGB}$ at the HL-LHC (a, b) and the CEPC (c, d). The black (blue) dashed curve denotes the contour of $F_{\rm PNGB}=0.7 ~(1.3)$, respectively. The red curve represents the $F_{\rm PNGB}=1$ contour with a 68\% C.L. error band. The green and yellow regions are excluded for $|\xi|\geq 0.5$. }
\label{fig1}
\end{figure}

Figure~\ref{fig1} displays the sensitivity to $F_{\rm PNGB}$ from the $R$ and $\Delta\kappa_{Z\gamma}$ measurements at the HL-LHC (a, b) and the CEPC (c, d). The latest global fit results of Higgs boson couplings show that the $\xi$ value is highly constrained~\cite{Khachatryan:2016vau,ATLAS-CONF-2018-031}. As both the Higgs nonlinearity and heavy new resonances can contribute to Higgs coupling deviations, these two effects might accidentally cancel each other out.  We choose three benchmark values of $F_{\rm PNGB}$'s for illustration; $F_{\rm PNGB}=0.7$ (black-dashed curve) and $F_{\rm PNGB}=1.3$ (blue-dashed) describes a PNGB Higgs boson from a compact and non-compact UV group, respectively, while $F_{\rm PNGB}=1.0$ (red-solid) denotes a SM-like Higgs boson. The shade band along each $F_{\rm PNGB}=1.0$ curve represents the 68\% C.L. uncertainty of the $F_{\rm PNGB}$ measurement, which is derived from Eq.~\ref{eq:errF1}. The black cross denotes a benchmark point of ($\Delta\kappa_{Z\gamma}$, $R$) and the corresponding errors given by Eqs.~\ref{eq:kza_error} and \ref{err_h}. 
The green and yellow shaded regions are not allowed as $|\xi|$ is too large ($|\xi|\geq 0.5$). 
Equipped with precision measurements of aTGCs and Higgs-boson couplings, one could determine $F_{\rm PNGB}=1.0\pm (0.15\sim 0.25)$ at the HL-LHC in the regions of $\Delta\kappa_{Z\gamma}\lesssim-0.03$ and $R\gtrsim 1$; see Fig.~\ref{fig1}(a). However, it is not possible to do so for a positive $\Delta\kappa_{Z\gamma}$; see Fig.~\ref{fig1}(b). 
In comparison with the HL-LHC, the uncertainty of aTGCs measurements at the CEPC is reduced by a factor of 10; see Eqs.~(\ref{eq:errF1}). As a result, the sensitivity to $F_{\rm PNGB}$ at the CEPC is improved greatly; for example, given a sizable $R$, one can determine $F_{\rm PNGB}=1.0\pm (0.03\sim 0.1)$ in the negative $\Delta\kappa_{Z\gamma}$ region and $F_{\rm PNGB}=1.0\pm  (0.05\sim 0.25)$ in the positive $\Delta\kappa_{Z\gamma}$ region. For a sizeable $\xi$, we could distinguish the Higgs boson nature since the accuracy of $F_{\rm PNGB}$ is much improved at the CEPC.

\noindent{\bf Conclusions.}
In this letter, we propose the signature of the PNGB Higgs boson can be distinguished in the ratio of $\mu(h\to VV^*)$ and $\mu(h\to Z\gamma)$ with the help of precision measurements of anomalous triple-gauge-boson couplings. The contribution of $O_H$ in the SM-EFT, which fakes the nonlinearity effect in $h\to VV^*$, is canceled out.  
Our result is valid in any coset $\mathcal{G}/\mathcal{H}$, as long as the unbroken group $\mathcal{H}$ in the IR contains the custodial $SO(4)$, of which the Higgs boson arises from a custodial $4$-plet. 
Depending on the magnitude of Higgs boson  nonlinearity parameter $\xi$, at least $1\sigma$ confidence level of experimental sensitivity can be reached in general with the prospected accuracy of $h\to Z\gamma$, $h\to V^*V$, aTGCs at HL-LHC and CEPC. Especially for the negative region of $\Delta\kappa_{Z\gamma}$, the sensitivity is good at the CEPC because aTGCs can be measured very accurately.

\noindent{\bf Acknowledgements.}
We thank Da Liu, Yandong Liu, Ian Low, Zhewei Yin, C.-P. Yuan, Chen Zhang, Hao Zhang for helpful conversations and Jiang-Hao Yu, Jue Zhang for comments on the manuscript.
This work is supported in part by the National Science Foundation of China under Grants No. 11635001, 11275009, 11675002, 11725520 and 11875072. BY is supported by the U.S. National Science Foundation under Grant No. PHY-1719914.

\bibliographystyle{apsrev}
\bibliography{reference}

\end{document}